\documentclass[global,twocolumn]{svjour}
\usepackage{graphicx,color,xspace,multirow,amssymb}
\journalname{Applied Physics B}
\def\fm#1{\ifmmode #1 \else $#1$\fi}
\def\ket#1{{%
  \ifmmode |\,#1\,\rangle \else $|\,#1\,\rangle$\fi}}
\def\bra#1{{%
  \ifmmode \langle\,#1\,| \else $\langle\,#1\,|$\fi}}

\def\nCa40{\fm{^{40}\mathrm{Ca}}\xspace}
\def\Ca40{\fm{\nCa40^{+}}\xspace}
\def\spz{\fm{{}^1\mathrm{P}_{0}}\xspace}
\def\dsoh{\fm{{}\ket{\mathrm{S}_{1/2}}}\xspace}
\def\dpoh{\fm{{}\ket{\mathrm{P}_{1/2}}}\xspace}
\def\ddth{\fm{{}\ket{\mathrm{D}_{3/2}}}\xspace}
\def\dpth{\fm{{}\ket{\mathrm{P}_{3/2}}}\xspace}

\def\dsohtdpoh{\dsoh\fm{\rightarrow}\dpoh\xspace}
\def\ddthtdpoh{\ddth\fm{\rightarrow}\dpoh\xspace}

\def\angL{\fm{{L}}\xspace}
\def\angJ{\fm{{J}}\xspace}
\def\mJ{\fm{{m_J}}\xspace}
\def\LJ{\fm{{L_J}}\xspace}

\def\dsohmJ#1{\fm{{}\ket{\mathrm{S}_{1/2},{#1}}}\xspace}
\def\dpohmJ#1{\fm{{}\ket{\mathrm{P}_{1/2},{#1}}}\xspace}
\def\ddthmJ#1{\fm{{}\ket{\mathrm{D}_{3/2},{#1}}}\xspace}

\def\mus{\fm{\mu\mathrm{s}}\xspace}

\begin{document}
\title{Raman spectroscopy of a single ion coupled to a high-finesse cavity}
\author{C.~Russo\inst{1}\thanks{authors contributed equally to this work} \and H.G.~Barros\inst{1,2\star} \and A.~Stute\inst{1,2} \and F.~Dubin\inst{1}\thanks{\emph{Present address:} Institute of Photonic Sciences (ICFO), 08860 Castelldefels (Barcelona), Spain} \and E.S.~Phillips\inst{1} \and T.~Monz\inst{1} \and T.E.~Northup\inst{1} \and C.~Becher\inst{1}\thanks{\emph{Present address:} Fachrichtung Technische Physik, Universit\"{a}t des Saarlandes, 66041 Saarbr\"{u}cken, Germany} \and T.~Salzburger\inst{3} \and H.~Ritsch\inst{3} \and P.O.~Schmidt$^1$ \and R.~Blatt\inst{1,2}
}                     
\institute{Institut f\"{u}r Experimentalphysik, Universit\"{a}t Innsbruck, 6020 Innsbruck, Austria \and Institut f\"{u}r Quantenoptik und Quantuminformation (IQOQI), 6020 Innsbruck, Austria \and Institut f\"{u}r Theoretische Physik, Universit\"{a}t Innsbruck, 6020 Innsbruck, Austria}
\mail{Piet O. Schmidt: Piet.Schmidt@uibk.ac.at}
\date{Received: date / Revised version: date}
%
\maketitle
\begin{abstract}
We describe an ion-based cavity-QED system in which the internal dynamics of an atom is coupled to the modes of an optical cavity by vacuum-stimulated Raman transitions. We observe Raman spectra for different excitation polarizations and find quantitative agreement with theoretical simulations. Residual motion of the ion introduces motional sidebands in the Raman spectrum and leads to ion delocalization. The system offers prospects for cavity-assisted resolved-sideband ground-state cooling and coherent manipulation of ions and photons.
\end{abstract}
%
\section{Introduction}\label{sec:intro}
The interaction of a single atom with a single optical mode is of considerable interest for a variety of applications in quantum computation and communication \cite{Pellizzari:1995a,Cirac1997,Briegel1998a,Zoller:2005a}.
Coherent interaction between a single atom and photons requires all loss mechanisms to be weaker than the coherent coupling. This is typically achieved by placing the atom inside a high-finesse cavity. Advances in controlling the internal and external degrees of freedom of the atom have lead to tremendous progress in neutral atom-based cavity-QED experiments. Several single photon sources \cite{McKeever:2004a,Hijlkema:2007a}, atom-mediated photon-photon entanglement \cite{Wilk:2007a}, and the reversible mapping of a specific photon state onto the internal state of an atom have been demonstrated \cite{Boozer2007}. Even better control over the internal and external atomic degrees of freedom can be achieved in trapped ion systems, as demonstrated by the achievements in quantum information processing \cite{Haeffner2005a,Leibfried2005,Reichle2006a,Benhelm:2008b,Blatt:2008}. This offers exciting prospects for ion-trap based cavity QED. Previous implementations have demonstrated that single ions can be trapped in electro-dynamic Paul traps for many hours, localized to a fraction of a wavelength within the standing wave of an optical cavity \cite{Guthohrlein:2001,Mundt2002}. Furthermore, coherent coupling between atom and cavity mode \cite{Mundt2002}, the generation of single photons on demand \cite{Keller:2004}, and the reduction of the excited state lifetime due to the presence of the cavity field \cite{Kreuter2004} have been shown. The implementation of cavity QED with trapped ions is technologically very challenging due to the cavity size constraints imposed by the ion trap. As a consequence, the strong coupling regime has not yet been reached.
Here, we report on a next-generation ion-based cavity-QED system (Section \ref{sec:setup}) in which we are able to drive Raman transitions between two atomic states while simultaneously emitting a photon into the cavity (Section \ref{sec:raman}). We investigate the spectra and polarization of the Raman resonances in the case of a trapped \Ca40 ion for different Raman drive laser polarizations (Section \ref{sec:results}). We find that simulations of the Raman spectra are in good agreement with the experiment. The experimental spectra exhibit motional sideband resonances of all three trap modes. This is a result of imperfect ion localization due to residual motion. A more quantitative understanding of the ion's localization with respect to the cavity field is gained by observing the visibility of the cavity's standing wave.

In this system, we are able to reach an intermediate coupling regime by adjusting the relative strength of coherent and incoherent processes to be approximately equal. This regime is characterized by intriguing, yet unexplored features that can be studied with our system.


\section{Setup}\label{sec:setup}
\begin{figure}
\begin{center}
    \includegraphics[width=4.5cm]{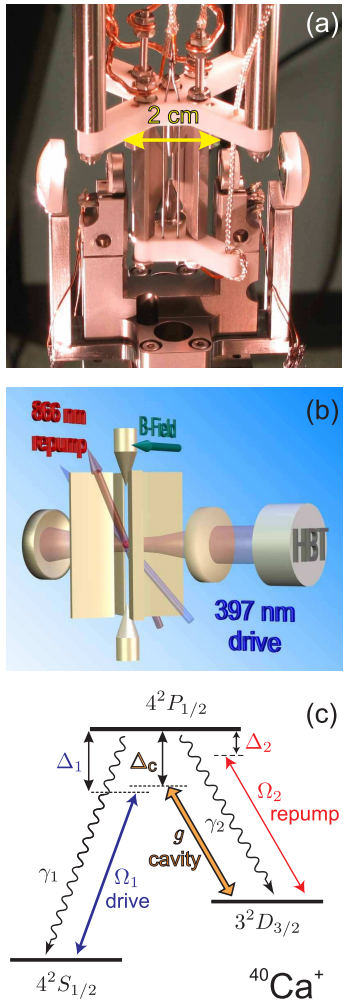}
    \caption{Experimental setup of the ion trap cavity-QED system. (a) Picture of the ion trap, surrounded by the cavity and two coupling lenses. (b) Geometric arrangement of the trap, cavity, magnetic and laser fields. HBT: Hanbury-Brown\&Twiss detection setup. (c) Simplified level scheme of \Ca40 showing the relevant ion-field couplings for driving Raman transitions and repumping. $\Omega_1$ and $\Omega_2$ are the Rabi frequencies of the drive and repump lasers, respectively, while $\Delta_1$ and $\Delta_2$ are their detunings from resonance. The spontaneous decay rates from the excited state are given by $2\gamma_1$ and $2\gamma_2$. The cavity is detuned by $\Delta_c$ from the atomic resonance and the ion-cavity interaction is characterized by the single-photon Rabi frequency $2g$.}
    \label{fig:setup}
\end{center}
\end{figure}
In the following we will briefly describe the experimental setup of our ion-based cavity-QED system. Further details can be found in \cite{Russo2008}.
\subsection{Ion Trap}\label{sec:trap}
A single \Ca40 ion is stored in a linear Paul trap \cite{Paul1990}, surrounded by an optical cavity as shown in Fig. \ref{fig:setup}(a). Radial confinement is achieved by applying a radio-fre\-quen\-cy of a few hundred volts at 23.4~MHz between two opposing pairs of razor-blade-shaped electrodes with a radial ion-electrode separation of 0.7~mm. Axial trapping is provided by applying a positive potential of typically 950~V to two tip electrodes opposing each other in the axial direction at a distance of 4~mm. The resulting trapping frequencies for a single \Ca40 ion are $\sim 3$~MHz and $\sim 1$~MHz in the radial and axial directions, respectively. Resistive heating of a metal tube generates a thermal beam of neutral \nCa40 atoms, which are then loaded into the trap via a two-step photoionization process: an external-cavity diode laser at 422~nm excites an electron to the \spz state of neutral calcium, from which it is ionized by a 375~nm laser. The ion lifetime in the trap typically ranges from several hours to days, allowing extended experimental investigations with one and the same atom.
\subsection{High-Finesse Cavity}\label{sec:cavity}
The ion trap is surrounded by two highly reflective mirrors, made by Research Electro-Optics (Boulder, CO, USA). The mirrors are separated by 19.92~mm in a near-concentric configuration which was determined by a spectral analysis of higher order cavity modes. The derived cavity waist is $13~\mu$m and gives a single-photon Rabi frequency of $2g=4\pi\times 1.61(2)$~MHz on the \ddthtdpoh transition (see Fig. \ref{fig:setup}(b)). However, this value is reduced to 1.4~MHz due to relative motion between ion and cavity (see Section \ref{sec:localization}).
The cavity features asymmetric mirror reflectivities, so that photons exit the cavity predominantly through one of the two mirrors. A cavity ring-down measurement yields a cavity finesse of $\approx 70(1)\times 10^3$ at 866~nm. The resulting field decay rate is $\kappa=2\pi\times 54(1)~$kHz. The properties of the mirrors have been carefully analyzed by measuring the reflected and transmitted power from both sides of the cavity, in an extension of the method described in \cite{Hood2001}. The transmission of the mirrors was measured to be 1.9(4)~ppm and 17(3)~ppm, whereas the total loss amounts to 71(3)~ppm. As a result, an intra-cavity photon is transmitted with 19(2)\% probability through the mirror with higher transmission.
The mirrors are mounted via piezo-electric transducers to a U-shaped mount which itself is mounted on a piezo-electric $xy$-stage that allows precise alignment of the cavity with respect to the ion. The cavity length is stabilized using a transfer lock scheme in which the cavity is locked to a diode laser at 785~nm. This laser is frequency-stabilized to an acoustically isolated and thermally stabilized high-finesse reference cavity \cite{Rohde2001a}, yielding a linewidth of $\approx 5$~kHz. We use the Pound-Drever-Hall method \cite{Drever1983} to derive an error signal from the optical resonance of this light with the trap cavity (finesse of $\approx 10000$ at 785~nm). A feedback signal is generated to adjust the length of one of the two piezos on which the mirrors are mounted. The system's first mechanical resonance at 5~kHz limits the bandwith of the feedback, resulting in residual frequency excursions of up to 350~kHz on a millisecond timescale.

\subsection{Laser systems}\label{sec:lasers}
Figure \ref{fig:setup}(b) shows a simplified level scheme of \Ca40. We use a frequency-stabilized and frequency-doubled Ti:Sap\-phire laser with a linewidth of $\approx 30$~kHz to address the \dsoh to \dpoh transition at 397~nm. This transition has a spontaneous decay rate $2\gamma_1=(2\pi\times 20.6~\mathrm{MHz})$. From the excited state, the atom can also decay at a rate of $2\gamma_2=(2\pi\times 1.69~\mathrm{MHz})$ to the \ddth state wich has a lifetime of 1.176~s \cite{Kreuter2005a}. Repumping back to the \dpoh state is achieved with a frequency-stabilized diode laser at 866~nm exhibiting a linewidth of $\approx 150$~kHz. Its polarization is set to drive $\sigma^+/\sigma^-$ ($\Delta \mJ=\pm 1$) transitions between the \ddth and \dpoh magnetic sublevels.
At 397~nm, we have a far-detuned laser beam to drive Raman transitions and a near-resonant Doppler cooling beam with a detuning of $\approx 10$~MHz red of the \dsohtdpoh transition. In the absence of the Doppler cooling laser, the Raman drive laser provides weak cooling of the ion via off-resonant scattering.
All experimental parameters, including Rabi frequencies, detunings and linewidths of the lasers, as well as the magnetic field, were calibrated by fitting dark resonance and excitation spectra, generated by sweeping the frequency of the repump laser with fixed detuning of the drive or cooling lasers. The typical errors are estimated to be below 10\% for all parameters.
\subsection{Detection system}\label{sec:detection}
\begin{table}[bt]
  \begin{center}
  \renewcommand{\arraystretch}{1.2}
  \begin{tabular}{|l|c|}
    \hline
    & transmission/\\
    \multirow{-2}*{optical element}  & efficiency
    \\\hline
    cavity output coupling & 19(2)\%\\
    fiber coupling and transmission & 92(2)\%\\
    filter and optical path & 67(2)\%\\
    APD1 & 42(2)\%\\
    APD2 & 41(2)\%\\\hline
  \end{tabular}
  \caption{Quantum efficiencies and transmission of the Hanbury-Brown\&Twiss setup. The combined error in the detection efficiency is estimated to be 14\%.}
  \label{table:detection}
  \end{center}
\end{table}
Fluorescence from the \dsohtdpoh transition is imaged onto a photo-multiplier tube (PMT, EMI P25PC) and an electron-multiplication charge-coupled device camera (EMCCD, Andor Ixon DV860AC-BV). We use narrow-band filters to block light other than 397~nm in both channels. The camera is used to visually confirm that only a single ion is trapped after loading. All other diagnostics and measurements on the \dsohtdpoh transition are performed using the PMT. The overall detection efficiency for fluorescence at the PMT is 1.7\% and the count rate from a single ion is $\approx 60$~kHz.

All measurements presented here are based on photons emitted by the high-finesse cavity. Photons leaving the cavity through the exit mirror are coupled into a multi-mode optical fiber. The fiber is connected to a Hanbury-Brown\&Twiss (HBT) setup consisting of a non-polarizing 50/50 beam splitter and two avalanche photodiodes (APDs, Perkin-Elmer SPCM-AQR-15). A Raman-edge filter at 830~nm removes the light from the transfer lock laser at 785~nm, whereas a narrow-band filter for 866~nm suppresses all other stray light. The dark count rates of the two APDs in the experiments presented here are $\sim 75$~counts/s and $\sim 180$~counts/s. The total background count rate of the APDs is $\sim 300$~counts/s for the experiments described in the following sections. A careful calibration of this system's detection efficiency was performed (see Table \ref{table:detection} for details). As a result, a combined count rate of both detectors of 32(4)~kcounts/s for a mean intra-cavity photon number of one can be inferred. Photo-detection events of the APDs are time-tagged with dedicated hardware (PicoQuant PicoHarp) with a timing resolution of 4~ps. The timing resolution of the APDs is 300~ps with a measured dead time of 88(4)~ns.
The HBT setup allows us to measure correlations between photons successively emitted by the cavity. However, for the experiments reported here, we sum the outputs of the two APDs, neglecting photon correlations.

\section{Vacuum-stimulated Raman transitions}\label{sec:raman}
While the experiments discussed here do not rely on coherence of the atom-cavity system, a primary objective of future research will be the study of coherent atom-photon interactions. This requires that all incoherent processes, such as cavity and atomic decay rates $\kappa$ and $\gamma=\gamma_1+\gamma_2$, respectively, are much smaller than the coherent interaction $g$. This so-called strong coupling regime is not met by our experimental parameters: $(g, \kappa, \gamma)=2\pi\times(1.61, 0.054, 11.1)$~MHz. However, this limitation can be mitigated by introducing a Raman coupling scheme in which the drive laser and cavity are detuned from the atomic resonance by $\Delta_1$ and $\Delta_c$, respectively (see Fig. \ref{fig:setup}(b)) \cite{Cohen-Tannoudji:1998}. In this situation the effective spontaneous decay is reduced, while maintaining a sufficiently high Raman coupling rate.
This intermediate coupling regime, in which all relevant rates are approximately equal, has interesting physical features that have not been explored in the past, such as the ability to accumulate photons in the cavity.

We drive vacuum-stimulated Raman transitions from \dsoh via \dpoh to the \ddth states by applying the drive laser at 397~nm detuned by $\Delta_1\approx 320$~MHz to the red of the \dsohtdpoh resonance. The cavity is at Raman resonance ($\Delta_c \approx 320$~MHz) with the drive laser (see Fig. \ref{fig:setup}(b)). We can adjust the resonance condition by either tuning the 397~nm drive laser frequency via acousto-optic modulators or tuning the cavity resonance via the 785~nm transfer laser.

\begin{table}
\centering
\begin{displaymath}
\begin{array}{|c|ccccc|}
\hline
 &  \textrm{from} &  \textrm{via virtual} &  \textrm{to} &   & \textrm{line}\\
\multirow{-2}*{\textrm{Line}} &  \dsohmJ{m_s} &  \dpohmJ{m_p} &  \ddthmJ{m_d} & \multirow{-2}*{$\frac{\Delta E_{sd}}{\mu_B B}$}  & \textrm{strength}\\
\hline
A &  +1/2 &  +1/2 &  -1/2 &  -7/5  & 1/18 \\
B &  +1/2 &  +1/2 &  +1/2 &  -3/5  & 1/9 \\
C &  +1/2 &  +1/2 &  +3/2 &  +1/5  & 1/6 \\
D &  -1/2 &  -1/2 &  -3/2 &  -1/5  & 1/6 \\
E &  -1/2 &  -1/2 &  -1/2 &  +3/5  & 1/9 \\
F &  -1/2 &  -1/2 &  +1/2 &  +7/5  & 1/18 \\
\hline
G &  +1/2 &  -1/2 &  -3/2 &  -11/5 & 1/3 \\
H &  +1/2 &  -1/2 &  -1/2 &  -7/5  & 2/9 \\
I &  +1/2 &  -1/2 &  +1/2 &  -3/5  & 1/9 \\
J &  -1/2 &  +1/2 &  -1/2 &  +3/5  & 1/9 \\
K &  -1/2 &  +1/2 &  +1/2 &  +7/5  & 2/9 \\
L &  -1/2 &  +1/2 &  +3/2 &  +11/5 & 1/3 \\
\hline
\end{array}
\end{displaymath}
\caption{Spectral lines associated with the vacuum-stimulated Raman transitions. Raman resonances from \dsoh to \ddth via a virtual \dpth state for different drive and vacuum-field polarizations are considered.
The effective line strength is calculated as the product of the Clebsch-Gordan coefficients associated with the drive and Stokes transitions composing the Raman transition.
See Fig. \ref{fig:pumpschemes}(c,d) for a graphical representation of these lines, including their individual coupling strengths, corrected for electric field projections as implemented in the experiment.}
\label{tab:ramanlines}
\end{table}

\begin{figure}
 \includegraphics[width=9cm]{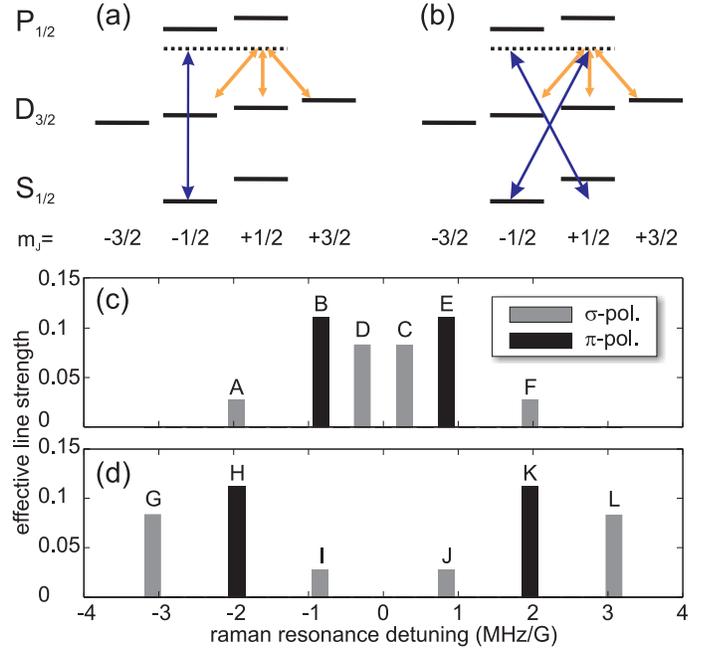}
 \caption{Driving schemes as used in the experiment and schematic representation of the corresponding spectra. (a), (b) Level scheme showing all eight electronic sublevels involved in the Raman transition for (a) $\pi$- and (b) $\sigma^+/\sigma^-$-polarized drive laser. As the drive laser is scanned, three Raman resonances can be addressed starting from a single \dsoh state. Effective line strengths and resonance frequencies of transitions for (c) $\pi$ and (d) $\sigma$ drive laser configuration: the linear polarization of the driving field is set parallel and perpendicular to the magnetic field, respectively. The height of the bars is proportional to the product of Clebsch-Gordan coefficients associated with the Raman transition times the projection of the electric fields. The light grey (black) bars indicate $\sigma$ ($\pi$) polarized photons emitted on the \ddthtdpoh transition, corresponding to vertically and horizontally polarized photons in the cavity. See Table \ref{tab:ramanlines} for more specific information on the lines A-L.
}
 \label{fig:pumpschemes}
\end{figure}

As shown in Fig. \ref{fig:pumpschemes}(a,b), up to eight electronic atomic levels contribute to the observed Raman spectra. In the following, atomic levels $\ket{\LJ, \mJ}$ are designated by their orbital angular momentum \angL, the total angular momentum \angJ and its projection \mJ onto the quantization axis which coincides with the magnetic field $B$. To quantitatively describe the system, numerical simulations using
the Quantum Optics Toolbox package \cite{Tan1999} for the MATLAB$^{\circledR}$ programming environment were developed \cite{Russo2008}. A master equation approach was used that includes all atomic levels and cavity excitations with a restricted Fock state basis, similar to \cite{Maurer2004}. In brief, we apply the dipole and rotating wave approximation in the atom-field interaction. The line strengths of the transitions and the projection of the field's polarization onto the quantization axis are taken into account for the lasers and the cavity field (see Fig. \ref{fig:setup}). The lifetime of the $D$-levels has been assumed infinite, which is a valid approximation considering the micro-second timescales of the atomic and cavity dynamics. Laser linewidths have been included by introducing appropriate collapse operators \cite{Gardiner:2004}. We neglect motion of the ion, which is a valid approximation only for a very strongly localized ion wave packet $z_0 k \ll 1$, where $z_0$ is the size of the wave packet of the ion in the ground state of the trap and $k$ is the wave number of the light field. The limitations of this simplification are discussed in Sections \ref{sec:spectrumcool} and \ref{sec:localization}.
A magnetic field $B$ shifts the state \ket{\LJ, \mJ} by an energy $\Delta E_{\LJ, \mJ}=-\mJ g_{\LJ} \mu_B B$, where $g_{\LJ}$ is the g-factor of the level $\ket{\LJ}$ and $\mu_B$ is Bohr's magneton.
Raman resonances can be assigned to transitions between magnetic sublevels \dsohmJ{m_S} and \ddthmJ{m_D} and experience a shift $\Delta E_{SD}=-(m_D g_D - m_S g_{S}) \mu_B B$. The effective line strength of a particular transition $\dsohmJ{m_S}\leftrightarrow\ddthmJ{m_D}$ via a polarization-selected state \dpohmJ{m_P} is calculated as the product of the Clebsch-Gordan coefficients for the corresponding $\dsohmJ{m_S}\leftrightarrow\dpohmJ{m_P}$ and $\dpohmJ{m_P}\leftrightarrow\ddthmJ{m_D}$ transitions.
Table \ref{tab:ramanlines} lists the magnetic field dependence, line strengths and involved sublevels of the Raman transitions for excitation with purely $\pi$, $\sigma^{+}$ or $\sigma^{-}$ polarized light and ideal mode-matching to the cavity field.

In our setup, the magnetic field is oriented perpendicularly to both, the cavity axis and the drive laser directions (see Fig. \ref{fig:setup}(a)). The linear polarization of the drive laser can be chosen to drive $\pi$ ($\Delta \mJ=0$) or $\sigma^+/\sigma^-$ ($\Delta \mJ=\pm 1$) transitions, which we call the $\pi$ and $\sigma$ drive laser configurations, respectively (see Fig. \ref{fig:pumpschemes}(a,b)). Since only one of the $\sigma^+/\sigma^-$ components of the drive field couples to the ion, the respective effective line strengths are reduced by half compared to the $\pi$ drive laser configuration. Similarly, the effective line strength of $\sigma$-polarized photons emitted into the cavity is half the maximum line strength. This stems from the non-vanishing electric field projections of the $\sigma^+/\sigma^-$ polarized Stokes photons onto the cavity axis. This part of the photon's field is not supported by the cavity mode, which results in a reduction of the observed line strength by a factor of two. Both effects lead to a reduction of the expected effective line strength and are accounted for in the effective line strengths shown in Fig. \ref{fig:pumpschemes}(c,d). The projection of the emitted photon's electric field onto the mode of the cavity field results in horizontally (vertically) polarized cavity photons for $\pi$ ($\sigma^{+}/\sigma^-$) transitions.

\section{Results}\label{sec:results}
\subsection{Raman spectra}\label{sec:spectrumpol}
\begin{figure}
    \includegraphics[width=9cm]{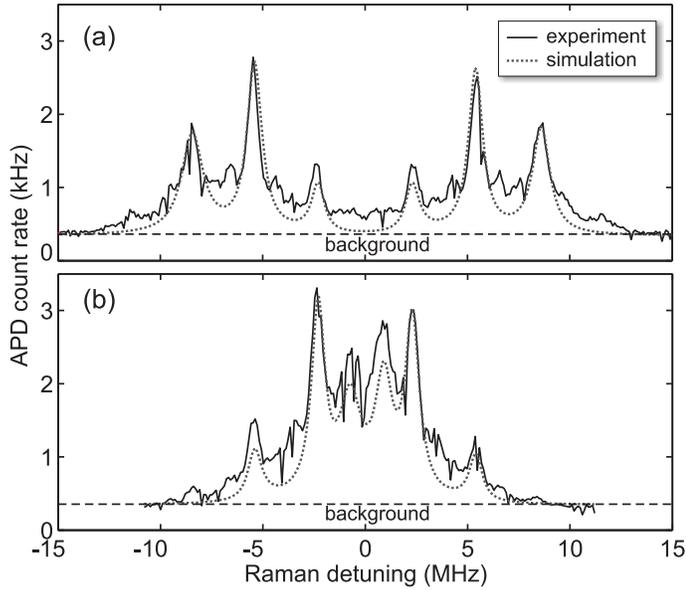}
    \caption{Raman spectra of the ion-cavity system for different drive laser configurations. Observed APD photon count rate as a function of the drive laser detuning for (a) $\sigma$ and (b) $\pi$ drive laser polarization. The dotted line is a simulation using independently calibrated parameters.}
    \label{fig:spectrumsigmapi}
\end{figure}
In a first experiment, Raman transitions in the ion-cavity system are observed for different drive laser polarizations. The drive and repump lasers are simultaneously applied with intensity-stabilized Rabi frequencies of $\Omega_1=82(8)$~MHz and $\Omega_2=7.7(7)$~MHz, respectively. The repump laser is detuned by $\Delta_2=0.5(2)$~MHz to the red of the \ddthtdpoh resonance. In addition, a near-resonant cooling laser with a detuning of $\approx$10~MHz below the \dsohtdpoh transition is applied to the ion to perform efficient Doppler cooling. As a function of the drive laser detuning from Raman resonance with the cavity, we count the number of photons detected by both APDs in a time interval of 1~s for each data point. The degeneracy between magnetic sublevels \mJ is lifted by a magnetic field of $B\approx$0.28~mT. Fig. \ref{fig:spectrumsigmapi}(a) and (b) show spectra of all six Raman resonances in the ion-cavity system for $\sigma$ and $\pi$ polarization of the drive laser, respectively.
A simulation of the Raman spectroscopy is shown as dotted lines in Fig. \ref{fig:spectrumsigmapi}(a,b). It is worthwhile to mention that the parameters entering the simulation are all independently calibrated (as described in Section \ref{sec:lasers}), but adjusted within their calibration error to match the spectra. The observed spectroscopic features are quantitatively reproduced by the simulation and agree with the effective line strengths and spectral separations as schematically depicted in Fig. \ref{fig:pumpschemes}(c,d).
\begin{figure}
    \includegraphics[width=9cm]{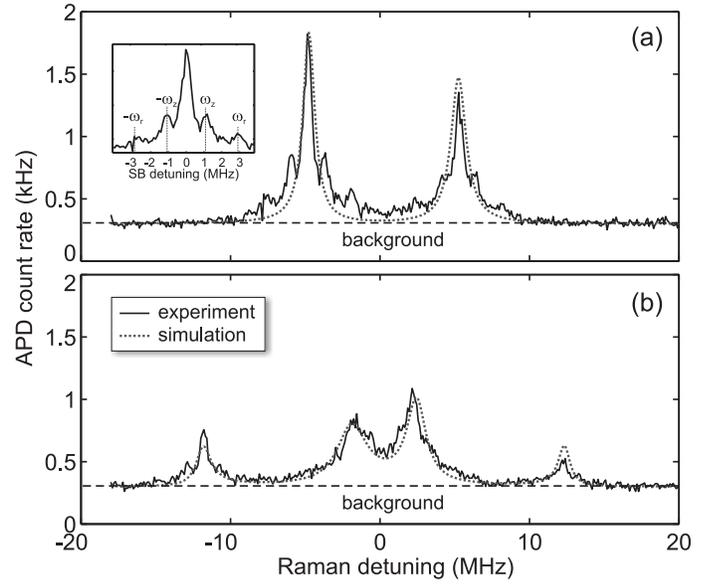}
    \caption{Raman spectra of the ion-cavity system for $\pi$-drive laser polarization. Observed APD photon count rate for (a) horizontally and (b) vertically polarized photons leaving the cavity as a function of the drive laser detuning. The dotted line is a simulation using independently calibrated parameters. The insert shows a zoom of the leftmost resonance indicating motional sideband resonances that are offset from the main resonance by $\pm 1.08$~MHz and $\pm 2.92$~MHz corresponding to the axial and radial trap frequencies $\omega_z$ and $\omega_r$, respectively.}
    \label{fig:spectrumpol}
\end{figure}

According to Table \ref{tab:ramanlines} and Fig. \ref{fig:pumpschemes}, peaks in the Raman spectrum correspond to either horizontally or vertically polarized cavity photons. This was confirmed by an experiment in which a polarizer was placed at the cavity output that allows us to selectively detect either horizontally or vertically polarized photons. The result of this measurement is shown in Fig. \ref{fig:spectrumpol} for $\pi$ drive laser configuration and a magnetic field of $B=0.61$~mT. Again, the simulation agrees quantitatively with the experimental data. The observed line strengths in Fig. \ref{fig:spectrumpol}(b) show an asymmetry which we attribute to preferential optical pumping into the \dsohmJ{+1/2} ground state by the repump laser. This is supported by the master equation simulation of the experiment. Small deviations in the strength of the resonances between theory and experiment are attributed to additional optical pumping by the near-resonant Doppler cooling laser, which is not accounted for in the simulations.
\subsection{Effect of motion on spectra}\label{sec:spectrumcool}
The experimental data in Fig. \ref{fig:spectrumpol}(a) shows additional resonances separated by approximately $\pm 1.08$~MHz and $\pm 2.92$~MHz from the main resonances with a reduced strength. These are identified with motional sidebands of the axial and radial motion of the ion in the trap \cite{Cirac1992}, at frequencies of $\sim 1~$MHz and $\sim 3~$MHz, respectively. This is a first proof-of-principle experiment showing that we can resolve and address all three oscillation modes of the ion in the trap. In future experiments, these well-resolved resonances could be used to cool the ion into the motional ground state, with the photon decay rate from the cavity being the dominant channel for dissipation \cite{Domokos2003,Zippilli2005}.

\begin{figure}
    \includegraphics[width=9cm]{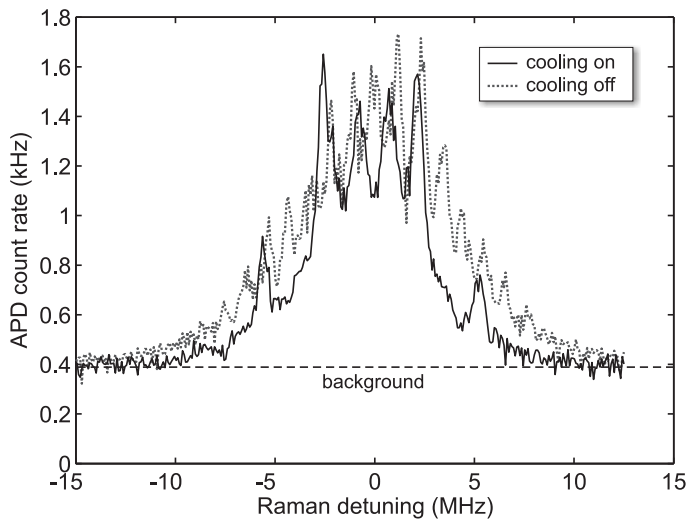}
    \caption{Raman spectrum of the ion-cavity system with motional sidebands. Observed APD photon count rate as a function of the drive laser detuning with (solid line) and without (dotted line) additional Doppler cooling beam.}
    \label{fig:spectrumcool}
\end{figure}
Since the sideband Rabi frequencies scale approximately as the square root of the number of motional excitations \cite{Wineland:1998a}, the effect of motional sidebands on the spectrum is expected to become more pronounced if the ion is hot.
Fig. \ref{fig:spectrumcool} shows Raman spectra in $\pi$ drive laser configuration with and without the Doppler cooling beam. The spectrum without Doppler cooling exhibits no clearly identifiable carrier resonances but rather a series of resonances including carrier and motional sidebands. We conclude that spectral information can only be obtained if the ion is sufficiently cold. This can either be achieved by continuously applying a Doppler cooling beam in addition to the Raman drive laser (as in Fig. \ref{fig:spectrumcool}), or in a pulsed excitation by interleaving Doppler cooling with spectroscopy.
\subsection{Ion localization}\label{sec:localization}
\begin{figure}
    \includegraphics[width=9cm]{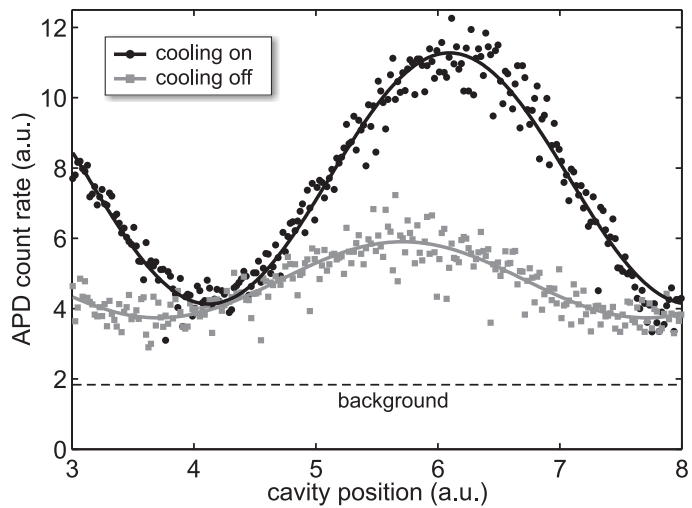}
    \caption{Standing wave of the cavity field observed by the ion. The black circles (gray squares) show the standing wave with (without) extra Doppler cooling applied to the ion. The solid lines are $\sin^2$ fits to the data.}
    \label{fig:standingwave}
\end{figure}
Another probe of the ion's temperature is its localization within the standing wave of the cavity \cite{Eschner:2003,Guthohrlein:2001,Keller2003}. For this, we tune the $\pi$-polarized drive laser to one of the Raman resonances and apply a 25~\mus drive laser pulse, followed by a 1~\mus wait time. After a 5~\mus repump pulse with the 866~nm laser, followed by another 1~\mus wait time, the sequence is repeated. During this last step, a near-resonant cooling laser can be applied to the ion for efficient Doppler cooling as described above. We have confirmed that this time is sufficient to reach steady-state conditions. During the whole sequence, we detect if a photon has been emitted by the cavity with the APDs. This way, typically 24000 cycles per data point are integrated. Changing the offset voltage of one of the two cavity piezos allows us to move the cavity field with respect to the ion for each data point. The other piezo is locked to stabilize the cavity length and follows the displacement of the first piezo. Fig. \ref{fig:standingwave} shows the result of the experiment with and without optional Doppler cooling. The standing wave visibilities obtained from the fits (solid lines in Fig. \ref{fig:standingwave}) are $V_\mathrm{cool}$=60\% and $V_\mathrm{nocool}$=35\% with and without Doppler cooling, respectively. The root-mean-square size $\sigma$ of a Gaussian wave packet describing the localization of the ion is related to the observed visibility by the relation $V=\exp\left(-2(k\sigma)^2\right)$, where $k=2\pi/(866~\mathrm{nm})$ is the wavenumber of the cavity field \cite{Eschner:2003,Keller2003}. This leads to a relative localization of the ion wave packet with respect to the cavity field of $\sigma_\mathrm{cool}=$70~nm and $\sigma_\mathrm{nocool}$=100~nm.
The comparably large ion delocalization in the absence of the Doppler cooling beam is an indication of insufficient cooling by the far-detuned drive laser. Although additional Doppler cooling improves the observed localization, the theoretical limit of $\sigma_\mathrm{theory}\approx 18~$nm is not reached.
We attribute this discrepancy to center-of-mass vibrations of the U-shaped cavity mirror mount. This is supported by an interferometric measurement of the cavity's motion.

Another consequence of the ion's delocalization is a reduced atom-cavity coupling strength $g$. From a localization of $70$~nm, we derive that $g$ is reduced from $g_\mathrm{max}=1.61$~MHz to $g_\mathrm{obs}=1.4$~MHz. This is confirmed by quantitative agreement between observed spectra and the numerical simulations, which use $g_\mathrm{obs}$ as one of the calibrated input parameters.

\section{Conclusion}
We have described a novel design for a cavity QED system based on trapped ions. We demonstrated that simulations of the Raman spectra using independently calibrated parameters obtained from dark resonance and excitation spectra can be used as a powerful tool to quantitatively predict the observed spectra.
Residual motion of the ion introduces clearly resolvable motional sidebands in the spectrum. This offers exciting prospects for motional ground state cooling of atomic or molecular ions using the loss of photons from the cavity as the dominant dissipative channel \cite{Domokos2003,Zippilli2005}.
Furthermore, adjusting the strength of Raman transitions allows us to tune the ratio between coherent and incoherent rates to place the system in an intermediate coupling regime. Future experiments will profit from the exquisite control over the internal and external degrees of freedom in ion trap experiments, combined with the ability to (coherently) couple an atom to the optical modes of a cavity.


\section{Acknowledgements}
We would like to thank C.~Monroe for stimulating discussions.
This work has been supported by the Austrian Science Fund (SFB 15), by the European Commission (CONQUEST Network, MRTN-CT-2003-505089; QUEST network HPRNCT-2000-00121; QUBITS network, IST-1999-13021; SCALA Integrated Project, Contract No. 015714); and by the "Institut f\"ur Quanteninformation GmbH". C. Russo acknowledges support from the Fun\-da\-\c{c}\~{a}o para a Ci\^{e}ncia e a Tecnologia -- SFRH/BD/6208/2001 and A. Stute acknowledges support from the Studien\-stiftung des deutschen Volkes.
\bibliographystyle{/TeX/bibliography/posplain2}
\bibliography{/TeX/bibliography/biblio-most-recent,/TeX/bibliography/addon}

\end{document}